# MD modeling of cracks in clay at the nanoscale

Zhe Zhang, Xiaoyu Song*

*Department of Civil and Coastal Engineering,
University of Florida, Gainesville, Florida*

**Abstract**

Cracks in clay are significant in geotechnical and geoenvironmental engineering (e.g., embankment erosion and stability of landfill cover systems). This article studies the mechanism of nucleation and growth of cracks in clay at the nanoscale through full-scale molecular dynamics simulations. The clay adopted is pyrophyllite, and the force field is CLAYFF. The crack formation in a pyrophyllite clay layer is evaluated under uniaxial tension and simple shear. The numerical results show that cracks in the nanoscale pyrophyllite clay layer are brittle and strain-rate dependent. Small strain rate results in low ultimate tensile/shear strength. As strain rate increases, clay crack shifts from a single-crack pattern to a multiple-crack one. The cracking mechanism is investigated from bond breakage analysis at the atomic scale. It is found that the first bond breakage occurs in the silicon-surface oxygen bond. As a crack propagates, the relative percentage of broken silicon-surface oxygen bonds is the smallest compared to other types of metal-oxygen interactions, demonstrating that the atomic interaction between silicon and surface oxygen atoms is the strongest. To understand the propagation of cracks, we also study the stress intensity factor and energy release rate of pyrophyllite and their size dependence at the atomic scale.

## 1. Introduction

In many geotechnical and geo-environmental engineering facilities such as embankments, dams, and landfill liners, crack in soils is an issue because it adversely affects the strength of soils (e.g., Miller et al., 1998; Alonso, 2021; Menon and Song, 2021). Crack has also been considered as a precursor to natural slope failures (Thusyanthan et al., 2007; Lu and Godt, 2013). The crack-induced strength loss matters because slope stability relies on the tensile strength of soil (Lu and Dong, 2017). In addition to affecting the mechanical properties of soils, cracks provide preferential flow paths for water, thereby enhancing the hydraulic conductivity compared to intact soils. The increase in hydraulic conductivity due to fracture could induce severe damages in the barrier system (Li et al., 2016; Lu and Kaya, 2013). Experiments have been carried out to investigate cracks in soil (Hanson et al., 1994; Hallett and Newson, 2001). Ayad et al. (1997) used field data from clay desiccation to predict the depth and spacing of primary shrinkage cracks. Zabala and Alonso (2011) used a strain-softening elastoplastic model to characterize the rupture process of the clay foundation. Kodikara et al. (2020) studied the mechanism of a desiccation-induced crack in thin soil layers. Tang et al. (2011) conducted laboratory tests to investigate the effect of wetting-drying cycles on crack development in clay layer using image processing. Lisjak et al. (2014) proposed a numerical approach based on non-linear fracture mechanics principles to gain insight into the failure mechanism of clay shales. Menon and Song (2019) studies desiccation cracking in unsaturated soils through a nonlocal numerical model. Cracking processes are macroscopic mechanical responses of a material that could be determined by constituent atoms and the associated laws of physics at the atomic level. However, most studies investigate cracks in clay at the continuum scale, ignoring the inherent nanoscale size of clay minerals. Clays usually

---

*Corresponding author

*Email address:* xysong@ufl.edu (Xiaoyu Song)

appear in the form of aggregates of tiny crystals of size smaller than 2 $\mu m$ (Hantal et al., 2014). Solid clay crystals large enough for direct laboratory testing are still rare (Wang et al., 2001). Because of their fine-grained polycrystalline nature, laboratory measurements of clay's structural and mechanical properties are challenging (Ortega et al., 2007; Likos et al., 2019). Moreover, cracks in clay could move discretely in the atomic lattice by breaking atomic bonds (Abraham et al., 1997).

Molecular dynamics simulations open new avenues in studying the nanoscale properties of material including clay (e.g., Teich-McGoldrick et al., 2012; Bitzek et al., 2015; Song et al., 2018; Song and Wang, 2019; Song and Zhang, 2021; Zhang and Song, 2021, 2022, among others). A propagating crack in a brittle material develops by breaking individual bonds between atoms (Bitzek et al., 2015). From an atomistic point of view, resistance to crack propagation should be characterized by the forces required to separate bonds successively at the crack tip. Chang (1970) first demonstrated that crack propagation is only possible on certain crystallographic planes and that crack tip bonds break after being stretched nearly to their elastic limit (Sinclair and Lawn, 1972). Atomistic dynamics dictate the complex pattern of crack nucleation and growth at the nanoscale. Molecular dynamics modeling provides a viable alternative to experimental techniques as it allows one to probe material properties on a molecular scale. There have been several numerical studies dedicated to the investigation of the anisotropic elastic properties of clays (e.g., Sato et al., 2005; Heinz et al., 2006; Zartman et al., 2010; Militzer et al., 2011; Ebrahimi et al., 2012; Teich-McGoldrick et al., 2012). In general, clay modeling requires an explicit treatment of atomistic dynamics (Abraham et al., 1997; Tjong, 2006). With the rapid development of supercomputers, molecular dynamics (MD) has become a viable method to investigate crack events at the microscopic scale (Abraham et al., 1994). MD allows revealing the full dynamical and atomistic nature of clay crack. However, these investigations have been largely qualitative, and most are limited in the description of mechanical properties. The study by Hantal et al. (2014) may be the first to determine the failure properties of clay minerals using a reactive molecule force field. They investigated different crack orientations. Clay layers have low fracture resistance under tensile loading perpendicular to the pre-defined crack, and the nanoscale crack mechanism is a stick-slip between clay layers under shear loading.

In clay minerals, the interlayer interactions are generally much weaker than the ion-covalent bonds in the tetrahedral and octahedral (T-O) layers. The weak cohesive bonding along the staking direction allows the rigid T-O layers to act almost as dislocation walls (Jia et al., 2021; Teich-McGoldrick et al., 2012). Jia et al. (2021) investigated the microscopic deformation and crack processes of kaolinite pores under high-pressure hydraulic fracturing water using MD simulations. Their numerical results indicated that the falling apart of tetrahedral or octahedral units results in cracks. Teich-McGoldrick et al. (2012) studied the elastic and structural properties of muscovite using MD simulations. They obtained mechanical properties as a function of temperature, pressure, and strain. Whereas there are multiple studies of clay's mechanical response using molecular dynamics, few studies have been devoted to understanding the crack mechanism. Through MD simulations, several studies (Duque-Redondo et al., 2014; Hantal et al., 2017) focused on the role of interfaces in elasticity and failure of clay and the failure mechanism of interlayer space of the clay platelet under different mechanical loading conditions. To the best of our knowledge, the nucleation and growth of cracks in a single clay layer under mechanical loading have been less studied.

This study is focused on understanding the atomic mechanism of the nucleation and growth of cracks in clay under tensile and shear loading. Also, it is worth noting that another motivation of this work is to provide a nanoscale physics foundation for modeling cracks in clay across various time and length scales through physics-based multiscale modeling techniques (e.g., Wagner and Liu, 2003; Xiao and Belytschko, 2004; Saether et al., 2009; Rafii-Tabar et al., 1998), which is beyond the scope of this article. Section 2 will introduce the molecular material model and numerical method. Section 3 presents the numerical results and discussions, followed by a closure.

## 2. Material model and numerical method

*2.1. Clay model*

The MD clay model consists of a single dry pyrophyllite layer in this work. The clay is flawless with no pre-existing crack. Pyrophyllite is a standard 2:1 structure clay type with a tetrahedral-octahedral-tetrahedral (T-O-T) structure (Wardle and Brindley, 1972). Its chemical formula is $Al_2[Si_4O_{10}](OH)_2$. The 2:1 structure forms charge-neutral layers where tetrahedral and octahedral layers are composed of Si and Al. In this study we select pyrophyllite for two reasons. First, it is the precursor to other clay minerals in the smectite group and provides mineral structure to these clays. Second, it has relatively high stability compared to other smectites such as montmorillonite which exhibits swelling. Since there are no Si and Al vacancies or metal substitution in its structure, pyrophyllite requires no interlayer cations for charge compensation. Thus, the neutral charge of the clay sheets inhibits swelling when interacting with water. For these reasons, we simulate the pyrophyllite clay layer to investigate its crack mechanism under tensile and shear deformation. Figure 1 shows the MD model of pyrophyllite. Atom type Ob, Si, and Al denote bridging oxygen, tetrahedral silicon, and octahedral aluminum. Ho and Oh are interlayer hydroxyl hydrogen and hydroxyl oxygen forming the hydrogen bond. Lattice parameters of crystal pyrophyllite are $a = 5.160$ Å, $b = 8.966$ Å, and $c = 9.347$ Å (Lee and Guggenheim, 1981). The crystal lattice is triclinic (non-orthogonal) with $\alpha = 91.18°$, $\beta = 100.46°$, and $\gamma = 89.64°$. We duplicate the unit cell of pyrophyllite to generate a supercell with $18 \times 10 \times 1$ units in the $a$, $b$, and $c$ dimensions, respectively. Hence, the initial dimensions of the pyrophyllite clay layer is 92.88 Å $\times$ 89.66 Å $\times$ 9.347 Å.

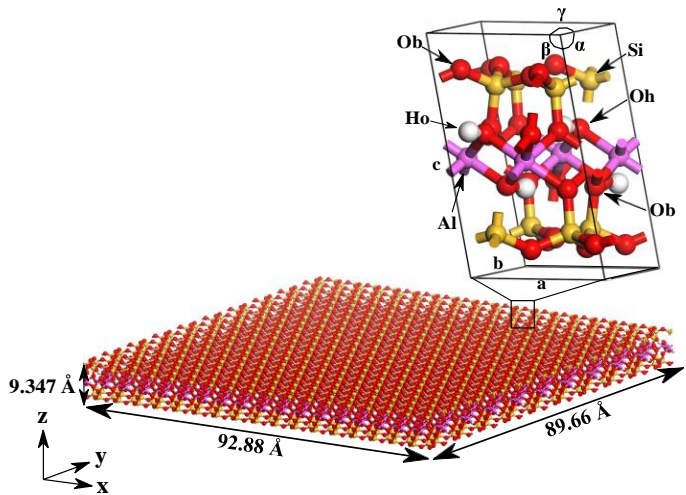

Figure 1: MD model of pyrophyllite.

## 2.2. MD Simulations

Molecular dynamics simulations are conducted using LAMMPS, a massively parallel molecular simulator (Plimpton, 1995). Atomic interactions are described through the CLAYFF force field (Cygan et al., 2004) with a cutoff radius $r_c$ equal to 10 Å. CLAYFF has been used to simulate deformation and crack processes of clay minerals (Fu et al., 2011; Hantal et al., 2014). The total potential energy $E_{total}$ consists of four terms, van der Waals energy described by Lennard-Jones (L-J) 12-6 potential, Coulomb electrostatic potentials calculated with partial charges, bond stretch energy, and angle bend energy described by the harmonic form. The PPPM (particle-particle particle-mesh) method (Hockney and Eastwood, 2021) is used to compute the long-range Coulombic interactions. The expression for each energy term can be found in Song and Zhang (2021). Periodic boundary conditions are applied in all directions. Temperature and pressure are maintained using Nose-Hoover algorithm (Hoover, 1985). Note that a Nose-Hoover thermostat will not work well for arbitrary values of the damping parameter. A good choice for temperature damping parameter and pressure damping parameter is of around 100 timesteps and 1000 timesteps, respectively (Plimpton, 1995). For this work, the time step is 1 fs (i.e., $10^{-15}$ s), and the damping parameters are 100 fs and 1000 fs for temperature and pressure, respectively.

The Verlet velocity algorithm is employed to integrate the equations of motion of atoms. We employ the isobaric-isothermal NPT ensemble (constant number of atoms, constant pressure, and constant Temperature) to equilibrate the atomic positions and volume at 298 K for 6 ns with a barostat coupling constant of 1000 fs. We couple the system to zero external pressure and allow the system to change the volume but not the shape. The equilibrium state is verified by the convergence of potential energy and system pressure. Figure 2 plots time variations of time-averaged potential energy and system pressure during equilibration. It indicates that the system has reached a steady state after 6 ns equilibration. The equilibrated system serves as the starting point for applying tensile or shear deformation. Since MD modeling is computationally expensive, all simulations are conducted with 128 CPU cores on the HiPerGator supercomputer. The wall clock time for the system to reach equilibrium for 6 nanoseconds simulation is about 6 hours. The elapsed run time for production at a tensile strain of 0.3 is around 3 hours.

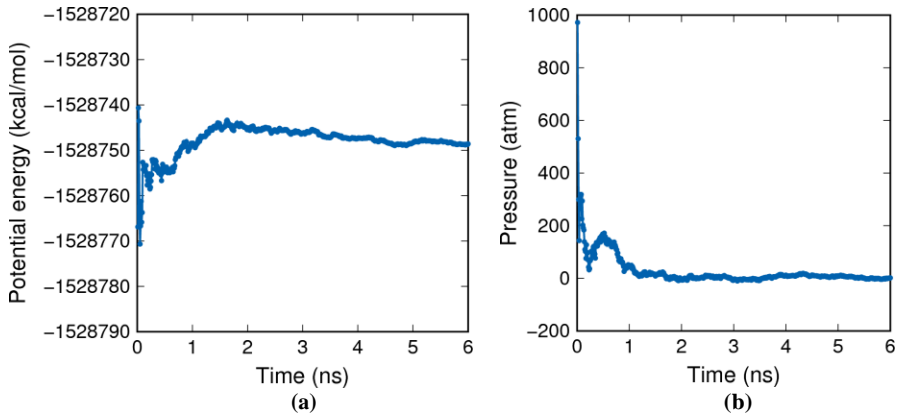

Figure 2: Time evolution of (a) potential energy and (b) system pressure during equilibration.

It is more convenient to control strain than stress in atomistic simulations. After the relaxation of the initial system, uniaxial tensile deformation is applied to the equilibrated clay particle by the strain-controlled loading method. The periodic simulation box is deformed at a constant engineering strain rate along a specified direction in this approach. At each time step when the dimension of the simulation box is changed, we remap atomic positions following the migration of the simulation box boundaries. The remap operation forces the atoms to deform via an affine transformation that exactly matches the deformation of the simulation box, which is appropriate for MD simulations of solids (Plimpton, 1995; Hantal et al., 2014). In the strain-controlled loading method, tensile strain is the ratio of length change to the original box length. As we deform the simulation box at a constant engineering strain rate, pressure components in the x, y, and z directions are controlled independently to allow transverse deformations due to Poisson's effect. For example, as the simulation volume is deformed along the x-axis, we apply the free transverse pressure condition by assigning zero pressure to the y and z directions of the system. The stress tensor is computed globally for all atoms in the system.

## 3. Numerical results

This section presents the numerical results of cracking in the flawless pyrophyllite clay layer from MD simulations. We first examine the global mechanical response of the system subject to tensile loading in the $x$- and y- directions. Here the x-direction and y-direction parallel the zigzag edge and armchair edge of the tetrahedral silicon layer in pyrophyllite, respectively. Then the atomic stress profile is presented to investigate material failure at the microscale. We report the nanoscale's crack propagation dynamics and fracture mechanisms from a bond breakage analysis. We study the fracture process at the nanoscale by calculating stress intensity factor and energy release rate during crack formation based on continuum fracture mechanics. A visualization tool, OVITO (Stukowski, 2009), is used to visualize and analyze atomistic simulation data in this study.

*3.1. Stress-strain curve under different loading rates*

This part presents the impact of loading rate on the stress-strain curve. It has been proven difficult to define stress tensor in a physically reasonable manner at the atomic scale in that stress is inherently a continuum concept. In this study we adopt the virial stress tensor to describe the stress state in clay The virial stress was developed based on the virial theorem by Clausius (1870) and Maxwell (1870). Virial stress was originally proposed to determine the stress field applied to the surface of a system containing atomic interactions (Zimmerman et al., 2004). Subramaniyan and Sun (2008) demonstrated the equivalence of the atomic scale virial stress and continuum scale Cauchy stress. Note that virial stress tensor could account for temperature effect by incorporating the contribution from kinetic energy. The varial stress tensor can be written as

$$\sigma_{ij} = \frac{1}{V}\sum_{A}\left[\frac{1}{2}\sum_{B=1}^{N}(r_i^B - r_i^A)F_j^{AB} - m^A v_i^A v_j^A\right], \tag{1}$$

where $V$ is the total volume of the atomistic system, $i$ and $j = x, y, z$ (i.e., the $x, y, z$ directions), $N$ is the number of neighboring atoms of atom $A$ within a cutoff radius $r_c$, $r_i^A$ and $r_i^B$ are the positions of atoms $A$ and $B$ along the direction $i$, $F_i^{AB}$ is the force exerted by atom $B$ on atom $A$, and $m^A$ and $v^A$ denote the mass and velocity of atom $A$, respectively.

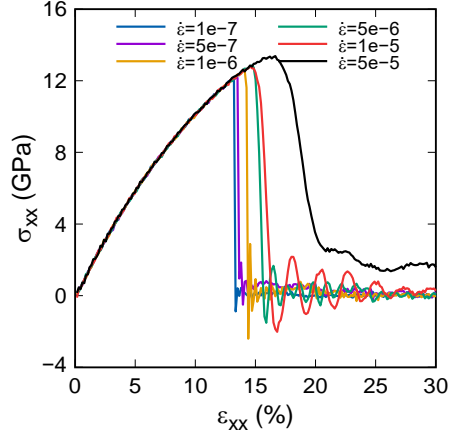

Figure 3: Stress-strain curves under different tensile loading rates in the x-direction.

Figure 3 shows the impact of strain rates on the stress-strain curves of pyrophyllite under uniaxial tension in the x-direction. The linear elastic regime corresponds to the range of strain from zero to 5%. The nonlinear stage takes place as the tensile deformation continues and strain hardening can be observed. Brittle fracture happens as the stress-strain curves abruptly drop to zero. The ultimate tensile strength exhibits the strain rate dependence. It increases with larger strain rate. For the case of the greatest strain rate simulated in this study, $\dot{\varepsilon} = 5 \times 10^{-5} fs^{-1}$, the ultimate tensile strength reaches 13.4 GPa at the strain around 16.7% while the lowest strain rate $\dot{\varepsilon} = 1 \times 10^{-7} fs^{-1}$ gives an ultimate strength of 12.2 GPa at the strain around 13.2%. Tensile stress begins to drop after the peak value as crack propagates. Once the tensile stress reaches zero, it starts to oscillate due to a spring-back phenomenon (Zhou et al., 2008). The strain rate below $5 \times 10^{-5} fs^{-1}$ has little to no effect on the stress-strain curve. Thus, in what follows we present the numerical results of simulations performed at a strain rate of $\dot{\varepsilon} = 1 \times 10^{-7} fs^{-1}$.

We also investigate the impact of the atomic structural anisotropy on the mechanical properties of clay under tensile deformation. Two loading scenarios are simulated. Tensile deformations are applied in the x direction for case 1 and in the y direction for case 2. Figure 4 compares the potential energy and stress-strain curves for two cases at the strain rate of $1 \times 10^{-7} fs^{-1}$. An earlier crack event occurs for case 2 at the tensile strain around 10.7% while for case 1 fracture occurs at strain around 13.3%. The ultimate tensile strengths for case 1 and case 2 are about 12.1 GPa and 11.6 GPa, respectively. This indicates that pyrophyllite has greater stiffness in x-direction than in y-direction, which is in agreement with the results in the literature (Teich-McGoldrick et al., 2012). Potential energy curve shows a higher peak value up to -1513508 kcal/mol for case 1 and a larger amount of energy release upon fracture event. By comparison, the peak value of potential energy for case 2 is -1516639 kcal/mol.

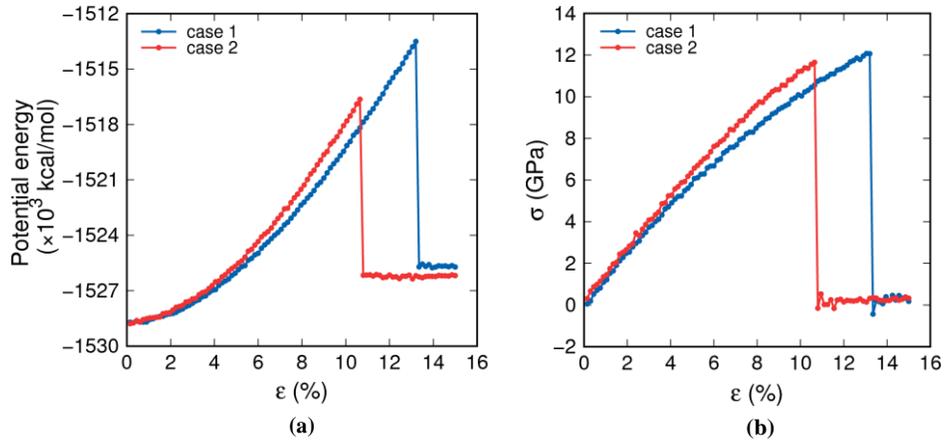

Figure 4: Effect of atomic structure anisotropy on (a) potential energy and (b) stress-strain curve under tensile loading. Note: case 1 and case 2 represents tensile strain applied in x- and y-direction, respectively. ($\dot{\varepsilon} = 1 \times 10^{-7} fs^{-1}$).

*3.2. Dynamics of crack nucleation and growth*

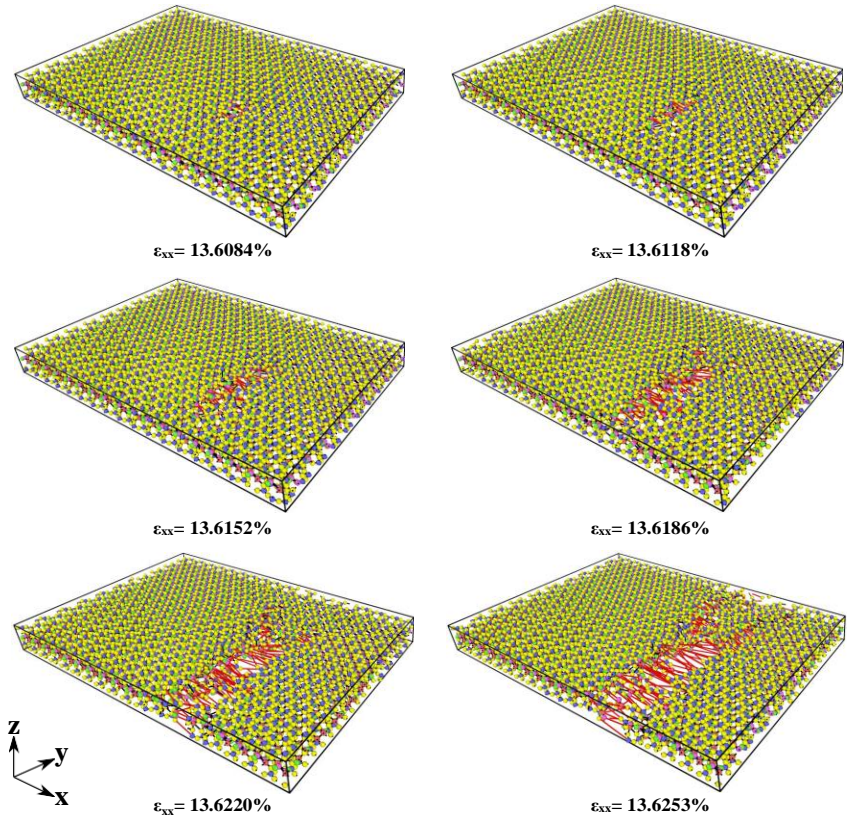

$\varepsilon_{xx}$= 13.6084%    $\varepsilon_{xx}$= 13.6118%

$\varepsilon_{xx}$= 13.6152%    $\varepsilon_{xx}$= 13.6186%

$\varepsilon_{xx}$= 13.6220%    $\varepsilon_{xx}$= 13.6253%

Figure 5: Crack nucleation and propagation under tensile load in x-direction with $\dot{\varepsilon} = 1 \times 10^{-7} fs^{-1}$.

Figure 5 presents the formation of cracks under tensile loading in x-direction with $\dot{\varepsilon} = 1 \times 10^{-7} fs^{-1}$. Crack develops perpendicular to the direction of the applied tensile load. Crack occurs first near the middle of the clay particle, and develops along both positive and negative directions of the y-axis. The increase in tensile strain for the crack from nucleation to full growth is only about 0.02%, indicating that the nanoscale crack in clay is brittle.

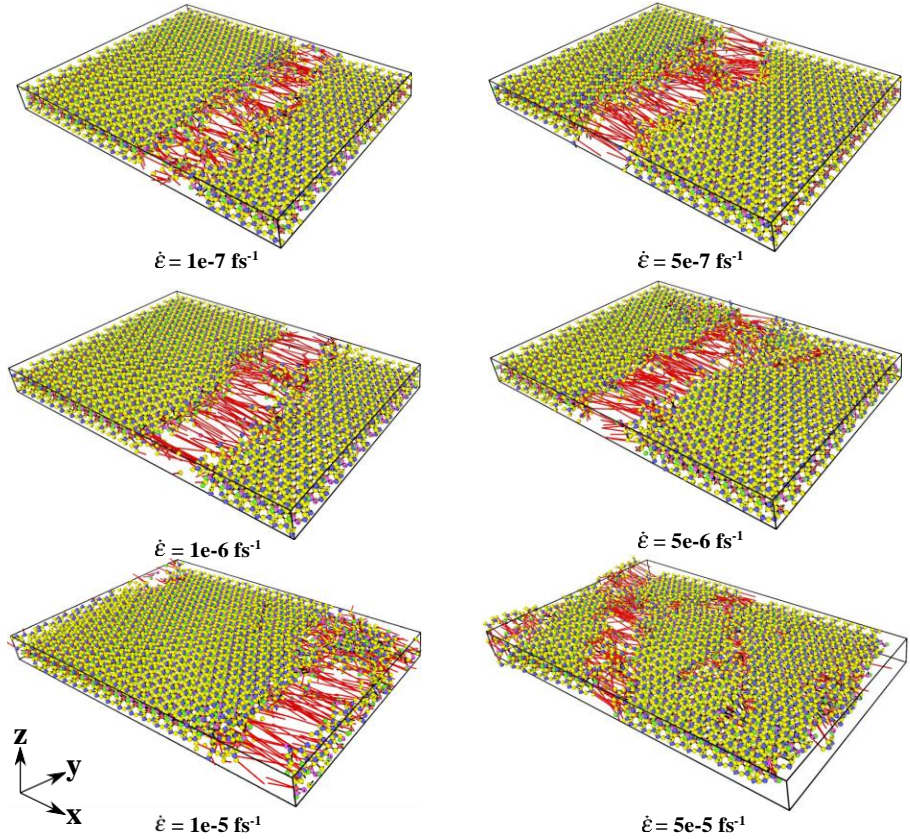

Figure 6: Comparison of crack patterns at different strain rates for tension in x-direction. ($\varepsilon_{xx}$ = 22.5%).

Figure 6 compares crack patterns at different strain rates ranging from $1 \times 10^{-7} fs^{-1}$ to $5 \times 10^{-5} fs^{-1}$. Bonds longer than the critical bond length are highlighted in red color to show crack surface. The determination of critical bond length to analyze bond breakage will be discussed in the next section. As strain rate increases, crack pattern switches from a single fracture to multiple slippage of rows of atoms. The difference is prominent between $1 \times 10^{-7} fs^{-1}$ and $5 \times 10^{-5} fs^{-1}$. Simulation results show that a strain rate less than $1 \times 10^{-6} fs^{-1}$ is sufficiently small to prevent the occurrence of multiple cracks.

We also investigate failure properties of pyrophyllite subject to simple shear deformation. In this case, shear strain is applied at a constant rate by shearing the periodic simulation box with the bottom boundary fixed. The shear stress-strain curve is plotted in Figure 7. The results show that the shear stress $\tau_{xy}$ increases until the shear strain reaches 14%. This could be caused by the elastic deformation of clay. After the shear stress reaches a maximum of about 9.8 GPa, it decreases significantly as shear strain further increases, signifying the clay has fractured.

We analyze the atomic stress profile in each layer because pyrophyllite is a T-O-T layered structure. Figure 8 shows the contour of atomic tensile stress field $\sigma_{xx}$ in (a) tetrahedral silicon layer, (b) surface bridging oxygen layer, and (c) octahedral aluminum layer at the tensile strain of 13.6%. Atomic stress is calculated using equation (1) where *V* becomes the atomic volume estimated by Voronoi method (Rycroft, 2009). In the silicon layer and aluminum layer, the minimum $\sigma_{xx}$ occurs near the inner boundary of crack. Further from the crack region, atomic stress increases. However, the stress concentration is not remarkable in these layers. $\sigma_{xx}$ is positive (i.e., tensile stress) in the silicon layer and aluminum layer. However, the surface oxygen layer sustains both tensile and compressive stress. This could be due to the different locations of oxygen atoms in the tetrahedral layer. Figure 9 shows the configuration of surface Si-Ob layer in the x-y and x-z plans. The results show that oxygen atoms with positive $\sigma_{xx}$ are always at location type 2 while oxygen atoms with negative $\sigma_{xx}$ are always at location type 1.

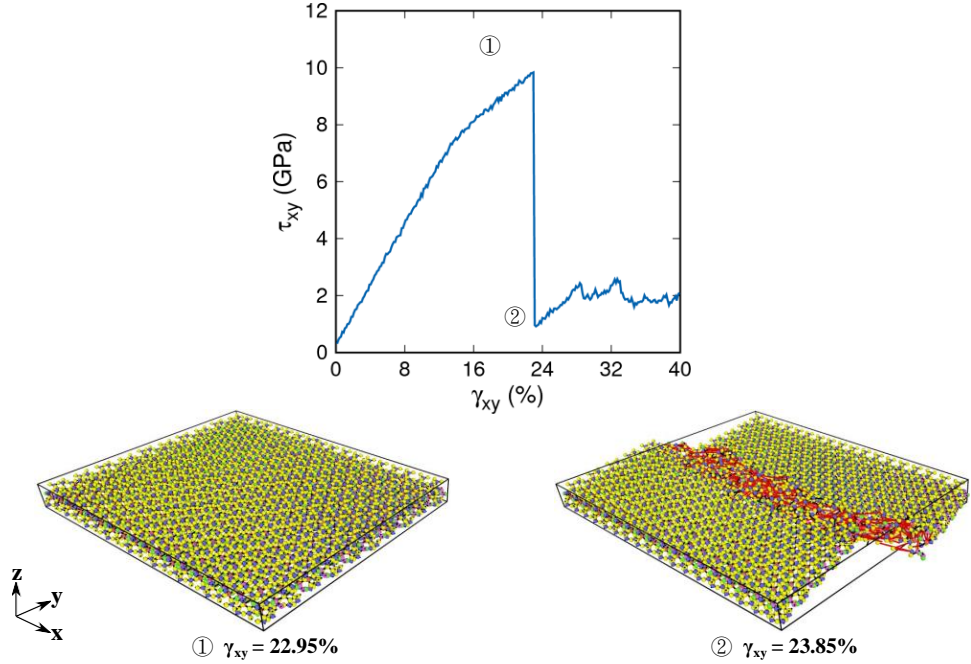

Figure 7: Stress-strain curve under simple shear with $\dot{\varepsilon} = 1 \times 10^{-7} fs^{-1}$.

Figure 10 compares the atomic velocity $v_x$ at four loading stages under tension in x-direction. Atoms on the left and right sides of crack surface moves in opposite directions at a similar rate. The results show little changes in velocity as crack propagates. Atomic velocities vary very slightly in the range of 0.0179 Å/fs to 0.0189 Å/fs for the atoms in the right side. This could be due to the tensile strain applied at a constant rate.

To quantitatively characterize the crack mechanism, we analyze bond breakage behaviors during crack propagation under two loading conditions, i.e., uniaxial tension and simple shear. For tensile deformation, we also consider the effect of molecular structural anisotropy of clay. The simulations are conducted with uniaxial tensile strain applied along x and y directions, respectively. The criteria for determining bond breaking is required to determine the formation of a crack in MD simulations. Rimsza et al. (2018) assumed that 2.25 Å could be a critical bond length at the onset of bond breakage for MD simulations of silica fracture. Chowdhury et al. (2019) found that the strained Si-O bond could be stretched up to 2.15 Å in silica. Note that their simulations employed a reactive force field that allows for continuous bond formation and breaking with appropriate bond order defined. Also, the material that they simulated is silica instead of clay minerals. For the reasons, the critical Si-O bond length cannot be directly applied to bond breaking analysis of pyrophyllite in this study. Alternatively, in this work we determine the critical bond length based on the potential energy curve and stress-strain curve. The critical loading time can be determined from Figure 3 and Figure 23, i.e., the timing of a sudden drop of the curves. Through a trial and error method, we find a bond length that matches the timing of crack occurrence. Prior to the critical loading time, bonds are stretched in the elastic range and no broken bonds can be observed. Based on multiple simulations, we determine the critical bond length as 5 Å. Thus, bonds longer than 5 Å are considered as broken bonds. Figure 11 shows bond connection at the tensile strain of 13.6242%. For visualization, we still keep broken bonds virtually connected but highlight them in red.

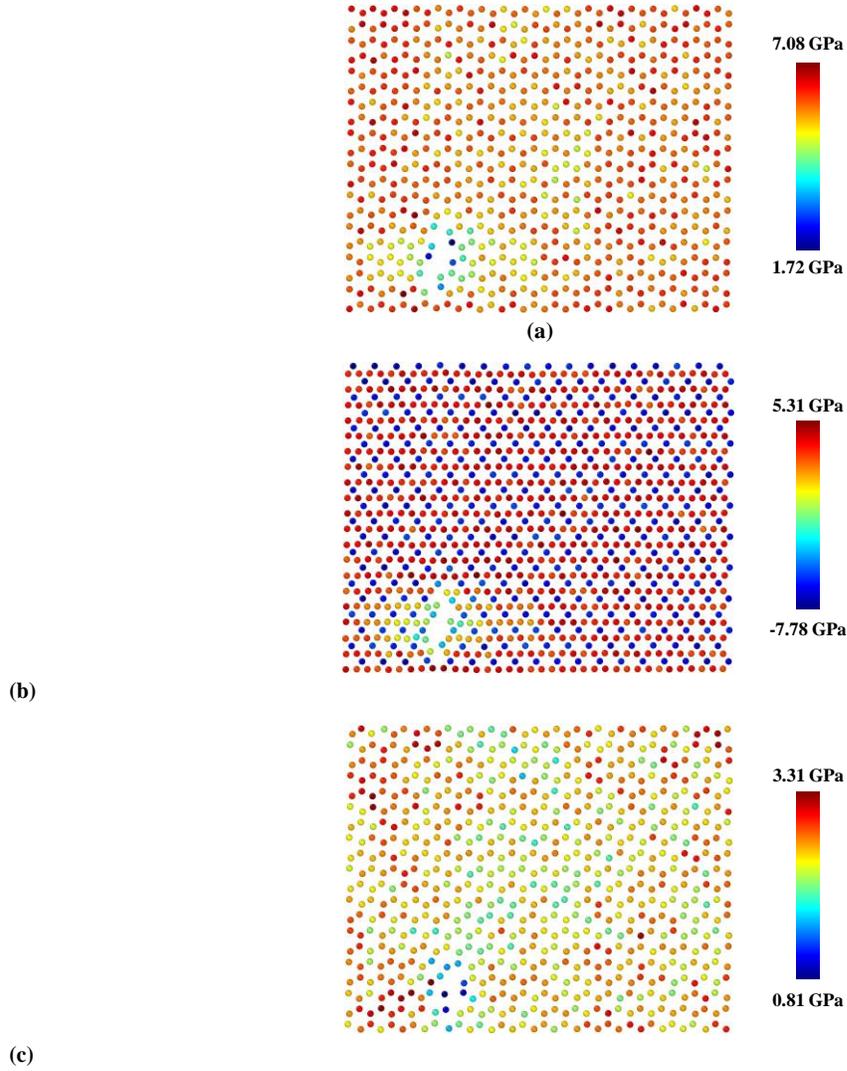

Figure 8: Contour of $\sigma_{xx}$ in (a) tetrahedral silicon layer, (b) surface bridging oxygen layer, and (c) octahedral aluminum layer at $\varepsilon_{xx} = 13.6\%$. ($\dot{\varepsilon} = 1 \times 10^{-7} fs^{-1}$).

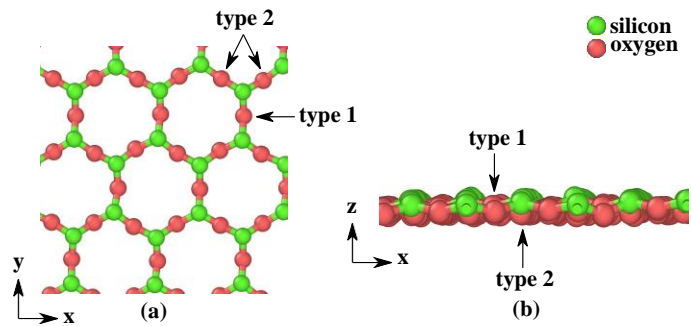

Figure 9: Configuration of surface Si-Ob layer in the (a) x-y plane and (b) x-z plane.

As elaborated earlier, the CLAYFF force field adopted in this study is based on a non-bonded description of the metal-oxygen interactions. In pyrophyllite, there are three types of metal-oxygen interactions, i.e.,

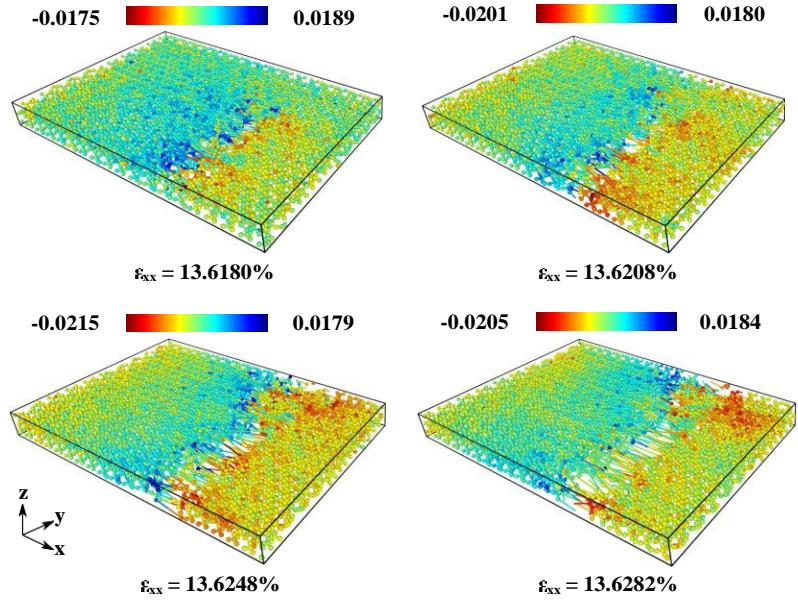

Figure 10: Contour of atomic velocity $v_x$ under tension in x direction with $\dot{\varepsilon} = 1 \times 10^{-7} fs^{-1}$. Velocity unit: Angstroms/femtosecond.

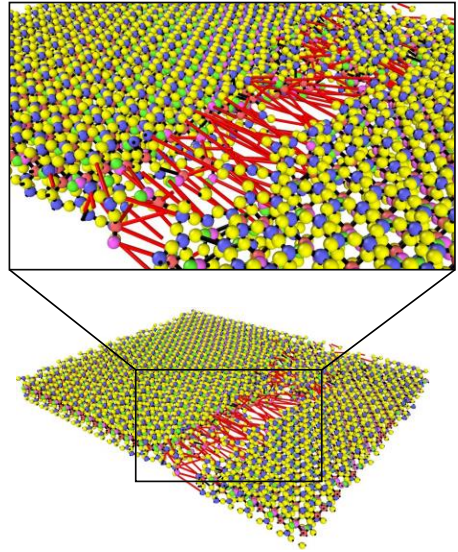

Figure 11: Visualization of bond connection under tensile strain in x-direction at $\varepsilon_{xx} = 13.6242\%$. Note: broken bond in red and unbroken bond in black.

Al-Oh, Al-Ob, and Si-Ob. The hydroxyl bond Oh-Ho is the only type of bonded interactions explicitly defined in CLAYFF. Each crack propagation is preceded by an extension of the Al-Oh, Al-Ob, and Si-Ob bonds around the crack tip. Figure 12 presents the direction of crack propagation in the silicon layer under tensile load in x direction. The strain rate is $1 \times 10^{-7} fs^{-1}$. The blue dot denotes the starting location of crack, i.e., the center of the first broken bond. The path of crack propagation is tracked based on the position of broken bonds that are plotted in red). Next, we study the number of broken bonds and percentage of each type of broken bonds in the nucleation and growth of the crack.

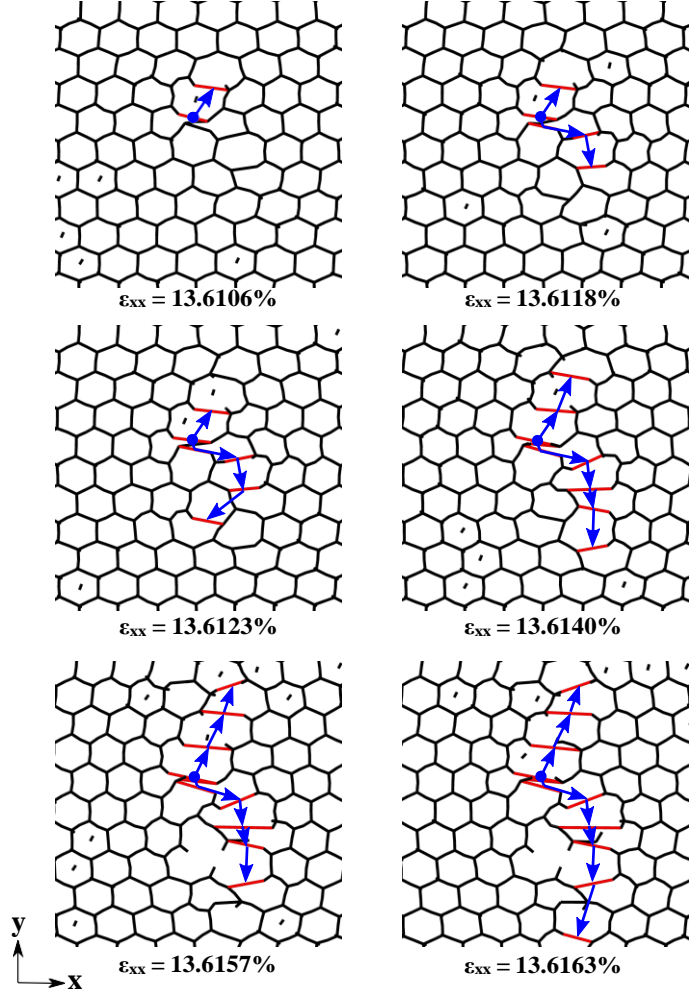

Figure 12: Direction of crack propagation in silicon layer under tension in x direction with $\dot{\varepsilon} = 1 \times 10^{-7} fs^{-1}$.

*3.2.1. Number of broken bonds*

Figure 13 shows the variation of the total number of broken bonds during crack growth. We consider the anisotropy effect by conducting two simulations with tensile strain applied in x and y directions, respectively. In both cases, the number of broken bonds first increases with crack nucleation and growth, and finally reaches an equilibrium indicating fully developed crack. The analysis of broken bonds helps to understand the effect of molecular structural anisotropy on crack formation in pyrophyllite at the atomic level. As shown in Figure 4(b), the ultimate tensile strength in x direction is greater than that in y direction. This is consistent with the broken bonds profiles. According to Figure 13, crack propagation under tension in x-direction results in 215 broken bonds while only 160 broken bonds are found for the case under tension in y direction. For the case under tension in y direction, the maximum number of broken bonds reaches at $\varepsilon_{yy} = 10.68\%$. However, for the case under tensile loading in x direction the maximum number of broken bonds occurs at $\varepsilon_{xx} = 13.63\%$.

*3.2.2. Percentage of broken bonds*

Three bond types Al-Oh, Al-Ob, and Si-Ob contribute to the bond breakage during crack propagation. The percentage of each type of broken bonds is shown in Figure 14. The percentage is defined as the ratio of number of one type of broken bonds to the total number of broken bonds. A bond starts to break in Si-Ob layer at the tensile strains of 13.604% and 10.67%, respectively, for the simulations under tensile

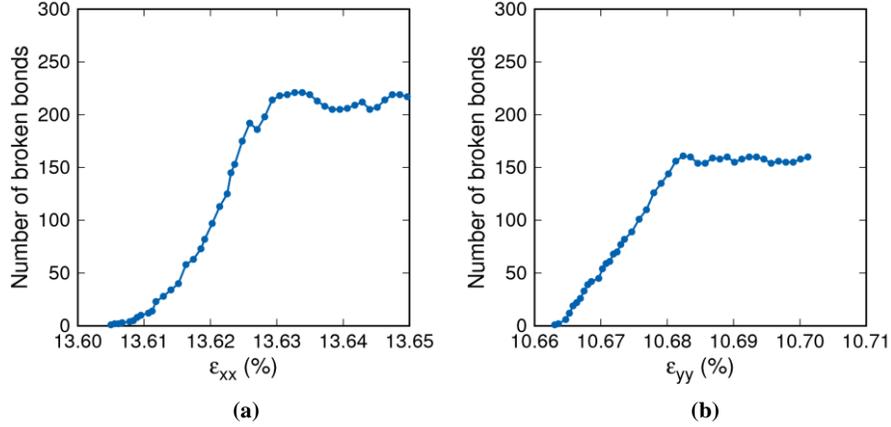

Figure 13: Variation of total number of broken bonds during crack propagation for tension in (a) x direction and (b) y direction ($\dot{\varepsilon} = 1 \times 10^{-7} fs^{-1}$).

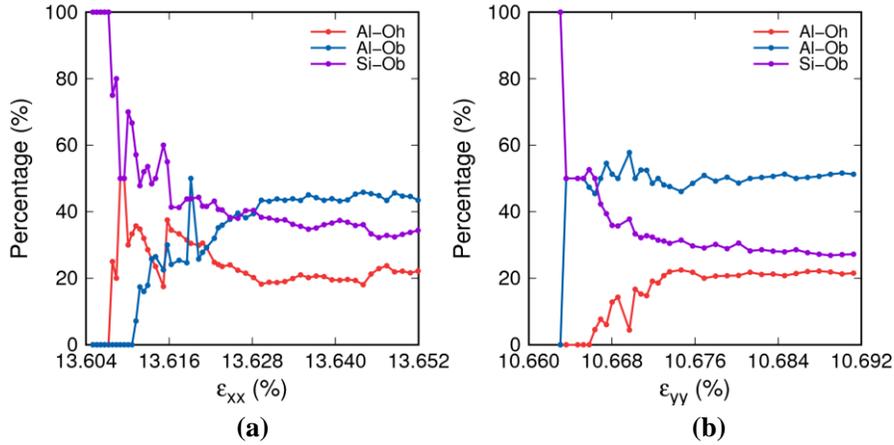

Figure 14: Percentage of broken bonds under tensile deformation along (a) x direction and (b) y direction. ($\dot{\varepsilon} = 1 \times 10^{-7} fs^{-1}$)

loading in the x- and y-directions. For both cases, Al-Ob bonds occupy the largest percentage of broken bonds, followed by Si-Ob and Al-Oh bonds in the order of strong to weak. For the tensile loading in x-direction, the final percentage of Al-Oh, Al-Ob, and Si-Ob broken bonds when crack is fully propagated are approximately 25%, 45%, and 35%, respectively. Similarly, for the tensile loading in y-direction the final percentage of Al-Oh, Al-Ob, and Si-Ob broken bonds are approximately 21%, 51%, and 28%, respectively. For both cases, no broken bonds are found in hydrogen bonds.

To explain the bond breakage behaviors of different types of bonds, we calculate potential energy of each bond type from van der Waals and Coulombic interactions. Figure 15 compares the potential energy between Al-Oh, Al-Ob, and Si-Ob. The results in Figure 15 show that the potential energy decay rate of Si-Ob interaction is slower than those of Al-Ob and Al-Oh. Al-Oh interaction decays slightly faster than Al-Ob interaction. This indicates that the Si-Ob interaction is stronger than the other bond types. It is noted that the percentage of broken bonds in Figure 14 is the absolute value, i.e., the number of one bond type divided by the total number of broken bonds. This value may not effectively distinguish the strength of different bonds. Alternatively, we plot the relative percentage which is defined as the ratio of the number of broken bonds to the total number of bonds in the MD system. Figure 16 demonstrates that Si-Ob is the strongest bond type, which is consistent with the nonbonded potential energy curve. For example, only 1.3% of total Si-ob bonds are stretched to break. By comparison, the relative percentage of Al-Oh and

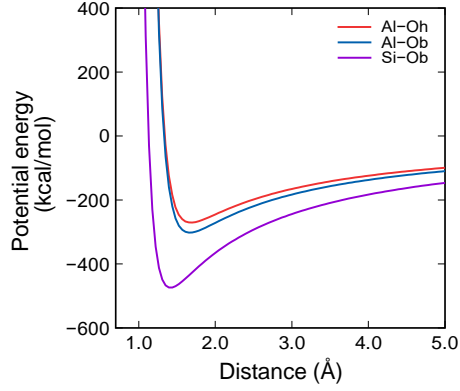

Figure 15: Nonbonded interactions between Al-Oh, Al-Ob, and Si-Ob.

Al-Ob bonds are up to 3.4%.

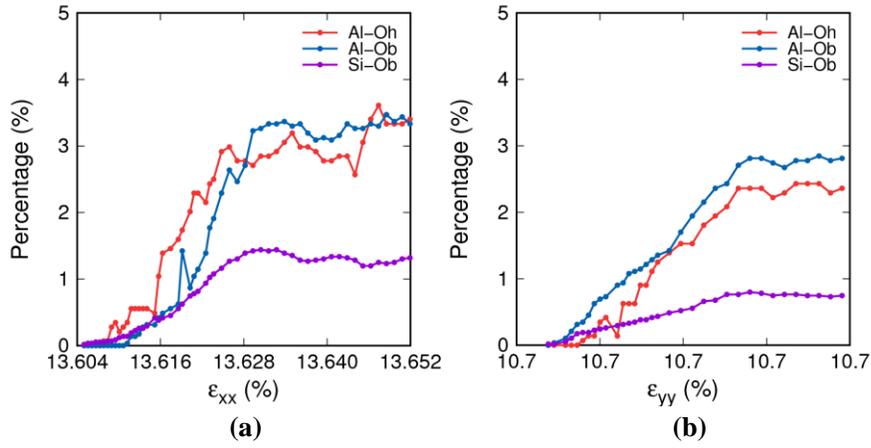

Figure 16: Relative percentage of broken bonds under tensile deformation in (a) x direction and (b) y direction at $\dot{\varepsilon} = 1 \times 10^{-7} fs^{-1}$.

Figure 17 shows the variation of the total number of broken bonds during crack propagation under simple shear deformation at strain rate of $1 \times 10^{-7} fs^{-1}$. Before $\gamma_{xy}$ reaches 22.29%, the number of broken bonds increases slowly. The majority of bond breakage events occurs between $\gamma_{xy}$ = 22.29% and $\gamma_{xy}$ = 22.305%. Little changes in broken bonds number are observed after $\gamma_{xy}$ = 22.31%. Figure 18 shows the absolute and relative percentage of each broken bond type during shear induced crack propagation. The minimum percentage is found for Si-Ob bond which proves its stronger interaction and resistance to bondbreakage.

### 3.3. Stress intensity factor

In continuum fracture mechanics the resistance of a material to crack propagation can be characterized by stress intensity factor (Anderson, 2017). For Mode I crack the stress intensity factor reads

$$K_I = \sigma\sqrt{\pi a}\left[\frac{W}{\pi a}\tan\left(\frac{\pi a}{W}\right)\right]^{1/2}, \quad (2)$$

where $\sigma$ is the tensile stress, $a$ is one half of crack length, $W$ is the width of the system. In this study, $W = L_x$ when tensile loading is applied in the y-direction. Figure 19 schematically illustrates the configuration for Mode 1 stress intensity factor $K_I$ calculation under tensile deformation along the y-direction. Note that the closed-form solutions (e.g,. equation (2)) for stress intensity factor were derived for continuum fracture mechanics. However, a few studies have successfully applied the closed-form solution to obtain $K_I$ from MD simulations (Le and Batra, 2016; Han et al., 2017). In this study we determine $K_I$ for Mode I crack in clay. The crack length is measured by the distance between two outermost broken bonds. Trajectories of atoms are output directly from MD simulations during crack propagation. Figure 20 plots the variation of crack length during crack propagation under tensile deformation. Both results of the simulations under tensile loads in the x-direction and y-direction respectively show approximately linear increase in crack length with respect to tensile strain. When tensile strain reaches at $\varepsilon_{xx}$ = 13.63% and $\varepsilon_{yy}$ = 10.68% respectively, the clay sample is completely fractured as indicated by a constant crack length. In general, for Mode I crack a critical value is referred to as the critical stress intensity factor, or fracture toughness of the material. That is, the stress intensity factor must exceed this critical value for crack to propagate in samples with a pre-existing crack. However, in this study no pre-crack exists in the clay sample before loading. Alternatively, we compute the maximum stress intensity factor for different loading conditions to compare the resistance to crack propagation. Figure 21 compares the variations of stress intensity factor during crack propagation at two strain rates, $\dot{\varepsilon} = 1 \times 10^{-7} fs^{-1}$ and $\dot{\varepsilon} = 5 \times 10^{-7} fs^{-1}$. Both cases give a similar value of the maximum $K_I$ around 1.05 MPa m$^{1/2}$. Note that strain rate below $5 \times 10^{-7} fs^{-1}$ also causes minor difference in the stress-strain curve (Figure 3). Figure 22 presents the effect of molecular structural anisotropy of clay on $K_I$. The anisotropy effect is remarkable that pyrophyllite has greater maximum stress intensity factor under tension in x direction than in y direction, with $K_I$ = 1.02 MPa m$^{1/2}$ versus $K_I$ = 0.9 MPa m$^{1/2}$. The greater fracture resistance of clay along x direction agrees with the greater ultimate strength in Figure 4 (b). The stress intensity factor obtained from MD simulations are in the same order with experimental data in the literature (Funatsu et al., 2004; Wang et al., 2007).

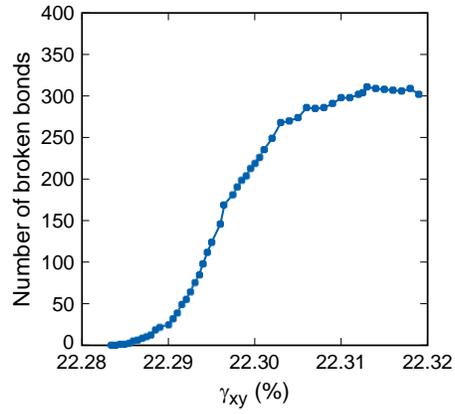

Figure 17: Total number of broken bonds under simple shear deformation at $\dot{\varepsilon} = 1 \times 10^{-7} fs^{-1}$.

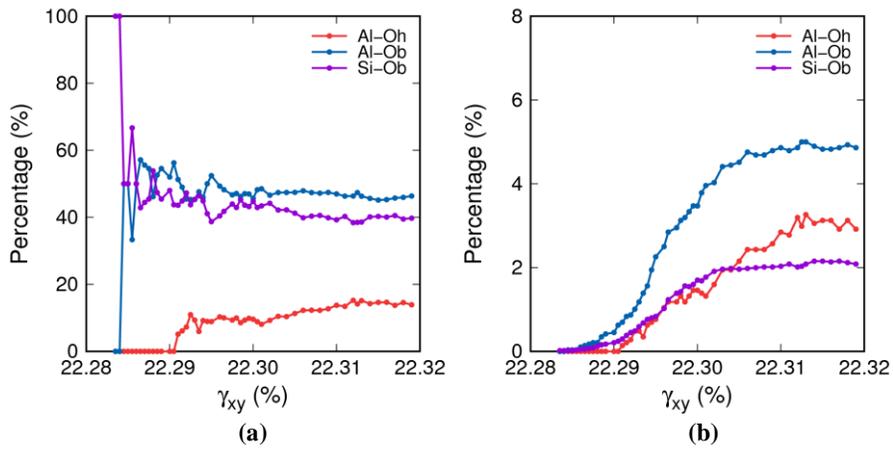

Figure 18: Absolute percentage (a) and relative percentage (b) of broken bonds under shear deformation at $\dot{\varepsilon} = 1 \times 10^{-7} fs^{-1}$.

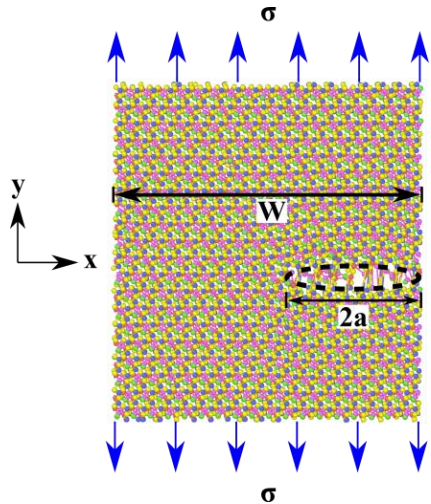

Figure 19: Schematic illustration of uniaxial tension and stress intensity factor calculation.

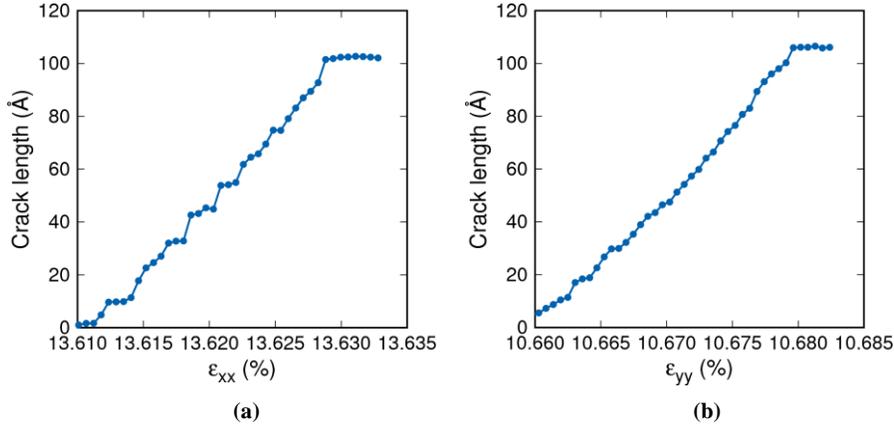

(a) (b)

Figure 20: Crack length during crack propagation under tensile deformation in (a) x-direction and (b) y-direction at $\dot{\varepsilon} = 1 \times 10^{-7} fs^{-1}$.

*3.4. Energy release rate of crack propagation*

In this part we study the energy release rate of crack propagation in clay at the nanoscale. Griffith (1921) proposed the Griffith energy balance for an incremental increase in crack area, $d\mathcal{A}$, under equilibrium conditions

$$\frac{d\mathcal{E}}{d\mathcal{A}} = \frac{d\mathcal{U}}{d\mathcal{A}} + \frac{d\mathcal{W}}{d\mathcal{A}} = 0, \quad (3)$$

where $\mathcal{E}$ is the total energy, $\mathcal{U}$ is the potential energy, $\mathcal{W}$ is the work required for crack growth, and $\mathcal{A}$ is the crack area. Irwin (1956) defined an energy release rate as the loss of total potential energy per incrementalincrease in crack area

$$G = -\frac{d\mathcal{U}}{d\mathcal{A}}. \quad (4)$$

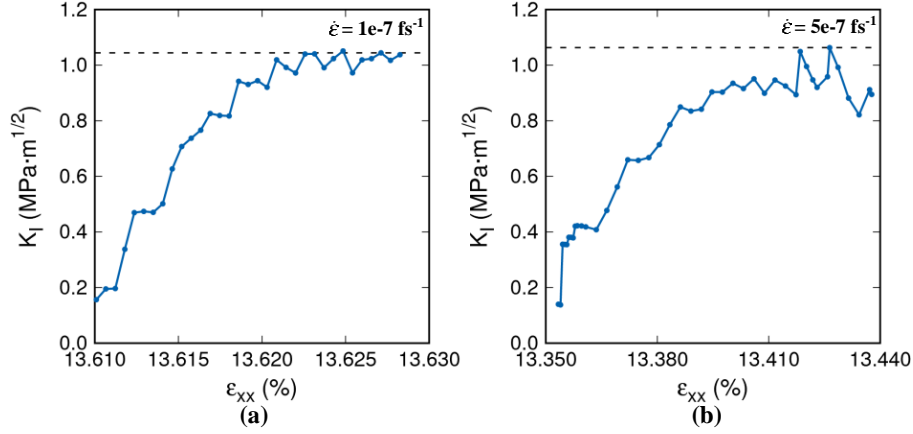

Figure 21: Stress intensity factor of clay under tension in x-direction at strain rates (a) $1 \times 10^{-7}$ fs$^{-1}$ and (b) $5 \times 10^{-7}$ fs$^{-1}$.

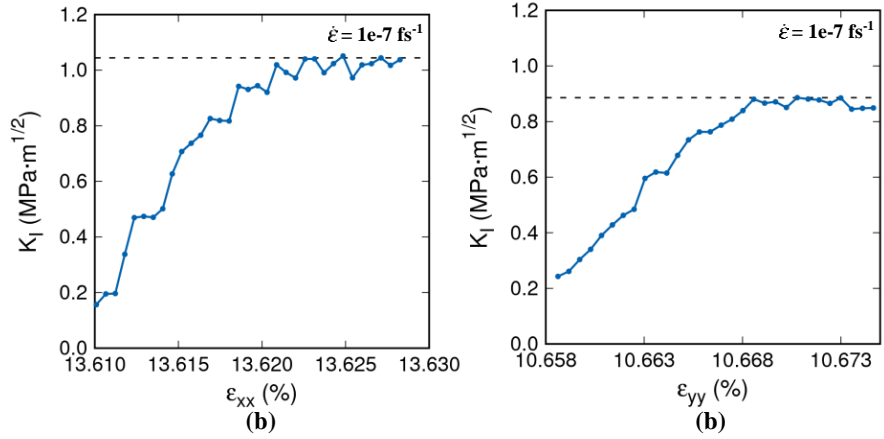

Figure 22: Stress intensity factor of clay under tension in (a) x-direction and (b) y-direction at strain rate $1 \times 10^{-7}$ fs$^{-1}$.

The energy-based approach for fracture propagation is essentially equivalent to the Griffith model. The term "rate" does not refer to the time derivative but the rate of change in potential energy with crack area (Anderson, 2017). The equations (3) and (4) have been successfully used to calculate energy release rate via MD simulations of several materials (e.g., Kikuchi et al., 2005; Jung et al., 2015; Liu et al., 2016; Bao et al., 2018). Thus, in this study we obtain energy release rate of pyrophyllite during crack propagation by equation (4). The potential energy is obtained from MD simulations and the crack surface area is determined through post-processing.

Figure 23 plots potential energy curves at different tensile strain rates. The results show that the smaller strain rate results in an earlier energy release and the larger strain rate causes a larger peak value of potential energy. The potential energy release curves at strain rates less than $5 \times 10^{-5} fs^{-1}$ exhibit a sharp drop while the energy release curve experiences a smooth decrease at strain rate $5 \times 10^{-5} fs^{-1}$. In what follows the analysis adopts the potential energy curve at strain rate $1 \times 10^{-7} fs^{-1}$. For simplicity, we assume an edge crack in pyrophyllite and the equation (4) can be written as

$$G = -\frac{d\mathcal{U}}{2\mathcal{T}\mathcal{L}_{crack}}. \tag{5}$$

where $\mathcal{T}$ is the thickness of pyrophyllite and $\mathcal{L}_{crack}$ is the crack length. The factor 2 stands for the two surfaces of the crack. Figure 24 shows the crack surface of the edge crack in clay at $\varepsilon_{xx} = 13.6242\%$.

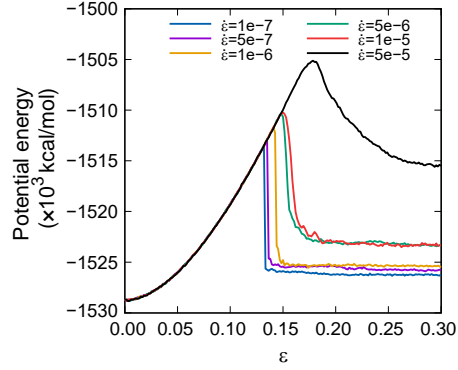

Figure 23: Effect of strain rates on potential energy curve under tensile deformation in x-direction.

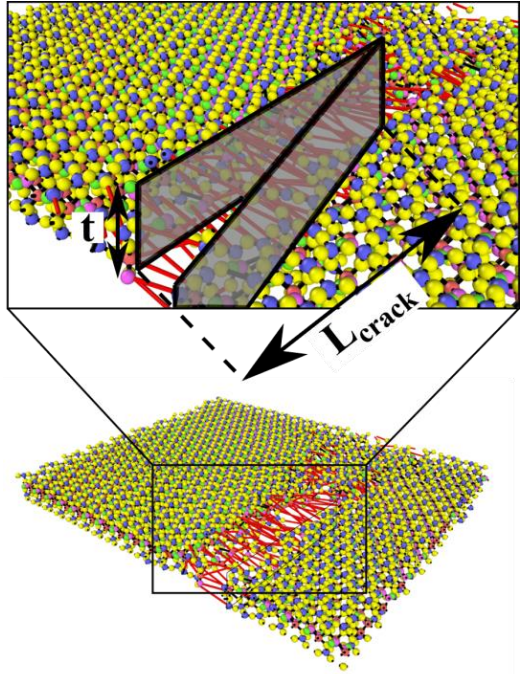

Figure 24: Schematic of the crack surface under tensile loading in x-direction at $\varepsilon_{xx} = 13.6242\%$.

Figure 25 shows the effect of strain rates on energy release rate during crack propagation under tensile loading in x-direction. Smaller strain rate results in an earlier energy release which indicates an earlier fracture event. As strain rate increases, we observe a larger peak value of potential energy. For example, energy release occurs at $\varepsilon_{xx} = 13.4\%$ and $\varepsilon_{xx} = 18.3\%$ at strain rate $1 \times 10^{-7} fs^{-1}$ and $5 \times 10^{-7} fs^{-1}$, respectively. Another difference between these curves is the rate of potential energy drop during crack growth. Smaller strain rate produces a sharp or abrupt drop during crack propagation while the potential energy dissipates more smoothly at greater strain rate. For example, the change in strain between maximum potential energy and equilibrium energy after the complete of crack is approximately 0.15% and 12% for strain rate $1 \times 10^{-7} fs^{-1}$ and $5 \times 10^{-7} fs^{-1}$, respectively. We calculate the change in potential energy between the initial configuration and the fully cracked clay at strain rate $1 \times 10^{-7} fs^{-1}$. For flawless clay particle, the change in crack surface area is equivalent to the area at the moment of fully cracked clay.

For the case of pre-existing crack, the critical energy release rate is calculated at the critical moment when the crack is about to propagate from its initial state. For flawless material in our work, however, energy release rate increases from zero with crack growth and finally reaches the dynamic equilibrium. Thus, we alternatively report the equilibrium energy release rate and compare them at different loading conditions. Figure 25 compares energy release rate at two different strain rates, $1 \times 10^{-7} fs^{-1}$ and $5 \times 10^{-7} fs^{-1}$ during crack propagation under tension in x-direction. However, minor strain rate effect on the equilibrium energy release rate is observed since both cases exhibit a dynamic value ranging between 5 to 6 *Joule/m²*. The finding that strain rate probably plays a negligible role in energy release rate has also been discussed in the literature for other materials (Chen et al., 2015; Liu et al., 2016). We also investigate the impact of molecular structural anisotropy anisotropy on energy release rate of pyrophyllite. Figure 26 shows the variation of *G* under tensile loading in x-direction and in y-direction, respectively, at the same strain rate $1 \times 10^{-7} fs^{-1}$. The equilibrium energy release rate *G* is about 4 *Joule/m²* under tensile loading in y-direction while *G* is about 5.5 *Joule/m²* for tensile loading in x-direction. This could indicate a larger fracture resistance of pyrophyllite in x-direction. The finding is consistent with the results in stress-strain curve and stress intensity factor considering effect of atomic structural anisotropy. The energy release rate calculated from our numerical simulations agrees with the experimental data in the literature (Yoshida andHallett, 2008).

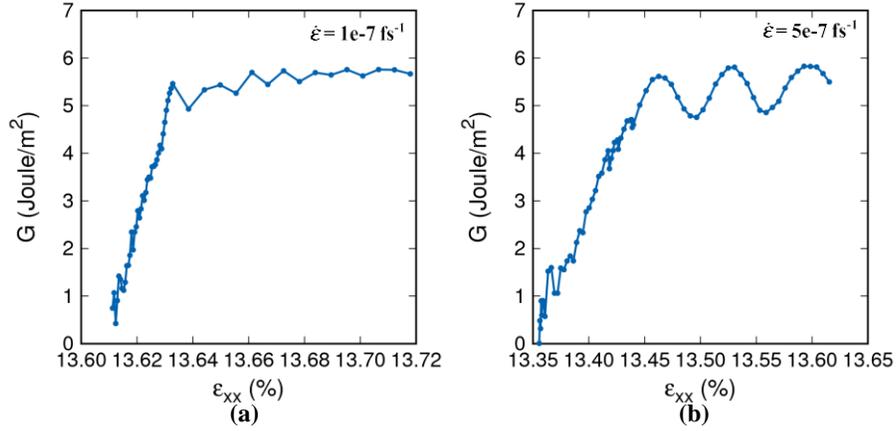

Figure 25: Energy release rate of crack in clay under tensile loading in x-direction at strain rates (a) $1 \times 10^{-7} fs^{-1}$ and (b) $5 \times 10^{-7} fs^{-1}$.

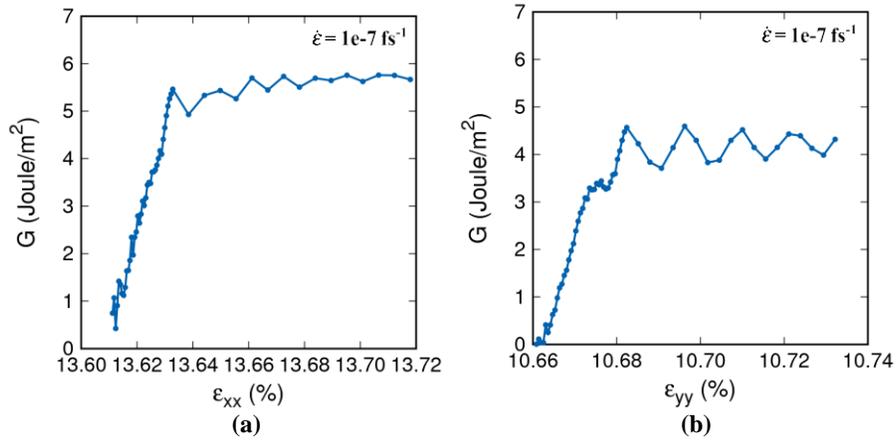

Figure 26: Energy release rate of crack in clay under tensile loading in (a) x-direction and (b) y-direction at the same strain rate $1 \times 10^{-7}$ fs$^{-1}$.

We briefly study the size effect on crack formation in clay. Three pyrophyllite models are constructed with $18 \times 10, 36 \times 20$, and $50 \times 29$ unit cells in the $x - y$ plane, respectively. Simulations are carried out under the same conditions, e.g., the timestep of 1 fs and tensile loading in x-direction, and a strain rate of $5 \times 10^{-7} fs^{-1}$. Figure 27 shows stress-strain curves of pyrophyllite consisting of different numbers of unitcells. The results show no noticeable size dependence of the stress strain curve.

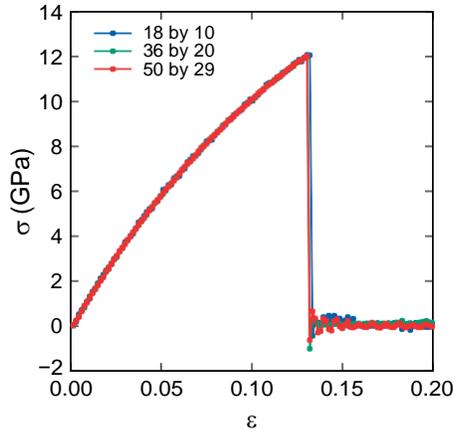

Figure 27: Size effect on stress-strain curves.

## 4. Closure

We have studied the mechanism of crack nucleation and growth in pyrophyllite through full-scale molecular dynamics simulations. The inception and propagation of cracks are modeled under uniaxial tension and simple shear conditions. The numerical results show that the pyrophyllite clay layer fractures in a brittle manner. The applied strain rate affects the mechanical response and crack patterns. Small strain rate results in low ultimate tensile/shear strength. As strain rate increases, clay fracture shifts from a single crack to multiple cracks. The fracture mechanism is investigated from bond breakage analysis at the atomic scale. We found that the first bond breakage occurs in the silicon-surface oxygen bond. As the crack propagates, the percentage of different types of broken bonds is compared. It is found that the relative percentage of broken silicon-surface oxygen bonds is the smallest indicating that the bond between silicon and surface oxygen atoms could be the strongest bond in clay. To understand the propagation of cracks, we also study the stress intensity factor and energy release rate of pyrophyllite and their size dependence at the atomic scale. These parameters are helpful to formulate a physics-based multiscale computational model for modeling cracks in clay at larger space and time scales.

**Acknowledgments**

This work has been supported by the US National Science Foundation under contract numbers 1659932 and 1944009.